\documentclass[aps,superscriptaddress,showpacs,twocolumn]{revtex4}
\usepackage{hyperref,amssymb,graphicx,bbm}

\begin{document}
\title{Exact invariant measures: How the strength of measure settles the intensity of chaos}

\author{Roberto Venegeroles}\email{roberto.venegeroles@ufabc.edu.br}
\address{Centro de Matem\'atica, Computa\c c\~ao e Cogni\c c\~ao, UFABC, 09210-170, Santo Andr\'e, SP, Brazil}

\date{\today}

\begin{abstract}
{The aim of this paper is to show how to extract dynamical behavior and ergodic properties from deterministic chaos with the assistance of exact invariant measures. On the one hand, we provide an approach to deal with the inverse problem of finding nonlinear interval maps from a given invariant measure. Then, we show how to identify ergodic properties by means of transitions along the phase space via exact measures. On the other hand, we discuss quantitatively how infinite measures imply maps having subexponential Lyapunov instability (weakly chaotic), as opposed to finite measure ergodic maps, which are fully chaotic. In addition, we provide general solutions of maps for which infinite invariant measures are exactly known throughout the interval (a demand from this field). Finally, we give a simple proof that infinite measure implies universal Mittag-Leffler statistics of observables, rather than narrow distributions typically observed in finite measure ergodic maps.}
\end{abstract}


\maketitle

\section{Introduction}
\label{sec1}

In many investigations we observe information that is generated by a physical system without prior knowledge of its nature. If such a system exhibits random-like behavior, then we will be interested in constructing the probability density function from its distribution of data. It is well-known that even one-dimensional deterministic dynamical systems in the form $x_{t+1}=T(x_{t})$ can exhibit random-like dynamical behavior. In the case of so-called measuring-preserving maps, the density of trajectories at time $t$, i.e., $\rho_{t}(x)$, has a peculiarity as $t\rightarrow\infty$: if we iterate randomly chosen initial points, the iterates will be distributed according to a limit density function $\rho(x)$ almost surely. In other words, $\rho_{t}(x)\rightarrow\rho(x)$, where $\rho(x)$ is the so-called invariant density of $T$, which satisfies
\begin{equation}
\label{invd}
\int_{T^{-1}(A)}\rho(x)dx=\int_{A}\rho(x)dx
\end{equation}
for any measurable subset $A$ of the phase space. The solution of Eq. (\ref{invd}) is the fixed point solution of the Perron-Frobenius operator \cite{BGB}
\begin{equation}
\label{PF}
\rho(x)=\sum_{\xi_{j}=T_{j}^{-1}(x)}\frac{\rho(\xi_{j})}{|T'(\xi_{j})|},
\end{equation}
where $T'$ is the derivative of $T$ and the sum extends over all preimages $T_{j}^{-1}$ of the point $x$ at which the density is to be evaluated.

The concept of invariant density has implications that go beyond theory of one-dimensional maps. We can mention applications in random number generators, dynamical models for oil drilling and for the Hipp pendulum regulator, Poncelet's closure theorem (a classical result of projective geometry), the Bogoyavlenskij-Novikov cosmological model (see Ref. \cite{BGB} for a review of all these problems), and ways to characterize patterns of activity in the olfactory bulb \cite{LLZ}, among other possibilities. As pointed out in Ref. \cite{CGST}, a further motivation for studying one-dimensional maps is that many higher-dimensional systems in the limit of strong dissipation approximately reduce to one-dimensional dynamics (Ref. \cite{HBL} provides a variety of experimental evidences).

Extracting statistical properties from deterministic systems based on long trajectories as data sets may not be a reliable procedure. It is possible that orbits display laminar or transient behavior for lengthy periods of time before settling into a fully chaotic mode. On the other hand, obtaining analytically an invariant density for a given map is typically a difficult task, when not unfeasible. The procedure to get a deterministic map equation from an invariant density given beforehand has proved more fruitful in the literature, and is known as the inverse Perron-Frobenius problem (IPFP) \cite{EM,DS,PSD,SK,BG,NC}. For numerical computations of the invariant density, the most used and best understood algorithm is Ulam's method \cite{Ulam,TLI}, a procedure for approximating invariant densities by using matrix approximations of the Perron-Frobenius operator [Eq. (\ref{PF}) is the unity eigenvalue case].

Here we develop a new approach to address the IPFP based on the manner whereby monotone branches of a map are related to each other, an idea recently proposed in Ref. \cite{TC} to deal with the conjugacy problem of maps. Beyond providing exact solutions of the IPFP, the aim of this paper is to outline the main ergodic differences between maps of finite and infinite (non-normalizable) invariant measures. The invariant measure $\mu$ is such that \begin{equation}
\label{dmu}
d\mu(x)=\rho(x)dx,
\end{equation}
i.e., it is assumed as absolutely continuous with respect to the Lebesgue measure \cite{nrd}, and will be the leitmotif for the results to be discussed here.

Note that $\mu$, when infinite, is not a {\it probability} measure. For maps having infinite invariant measure the conventional Lyapunov exponent vanishes, resulting in weakly chaotic behavior. Such kinds of maps are the motivation of so-called infinite ergodic theory \cite{Aaronson}, an issue that is gaining increasing interest in the literature \cite{GW,XWS,TPS,CG,Thaler,IGR,BGBL,Zw,SVp,RVT,SVPR,RK,RV1,NV}. Despite its sound development, this theory still suffers from a paucity of models whose invariant measures are exactly known, and this paper aims to bring further alternatives. It is also noteworthy that, as far as I know, numerical methods for estimating infinite invariant measures do not seem to be available or hinted at in the literature \cite{note}, which highlights the importance of exact results.

Exact invariant measures are also employed here to study the evolution of time averages of observables (like finite-time Lyapunov exponents) in a more efficient way. The approach, initially developed in Ref. \cite{NV}, is extended to more general shapes of maps, also enabling us to identify geometric characteristics of them. Based on this approach we also provide a simple proof that infinite measure implies universal Mittag-Leffler statistics of observables, also the so-called Aaronson-Darling-Kac theorem, a central result of infinite ergodic theory \cite{Aaronson}.

\section{Measure and conjugacy} 
\label{sec2}

Let us consider a Markov map $T:I\rightarrow I$ defined on a partition $\{I_{0}, \ldots, I_{k-1}\}$ of $I$ so that all of $k$ branches $T|_{I_{j}}=T_{j}$ are monotone. Thus, we can relate each branch $T_{j}$ with the absolutely continuous invariant measure of the map in the form
\begin{equation}
\label{adf}
T_{j}=\Phi_{j}^{-1}\circ\mu,
\end{equation}
where $\Phi_{j}$ are monotone functions. Note that each branch $T_{j}$ is also a function extending on the whole interval $I$ and with an image beyond $I$. Let us also introduce the foldings $h_{j}$, relating each branch of $T$ with $T_{0}$, i.e., $T_{j}=h_{j}\circ T_{0}$. Thus we have
\begin{equation}
\label{fol}
h_{j}=\Phi_{j}^{-1}\circ\Phi,
\end{equation}
where we simply set $\Phi_{0}=\Phi$ and $h_{0}(x)=x$. The corresponding invariant density satisfying Eq. (\ref{PF}) is such that
\begin{equation}
\label{PF2}
\rho=\frac{\rho\circ T_{0}^{-1}}{|T'_{0}\circ T_{0}^{-1}|}+\sum_{j=1}^{k-1}\frac{\mbox{sgn}(h'_{j})}{h_{j}'\circ h_{j}^{-1}}\frac{\rho\circ T_{0}^{-1}\circ h_{j}^{-1}}{|T'_{0}\circ T_{0}^{-1}\circ h_{j}^{-1}|},
\end{equation}
where $h'_{j}$ and $h_{j}^{-1}$ denote, respectively, the derivative and the inverse of $h_{j}$ (the same holds for $T_{0}$), and $\mbox{sgn}$ stands for the sign function. Within such an approach, the invariant measure can be directly calculated by integration of Eq. (\ref{PF2}), namely \cite{TC},
\begin{equation}
\label{muvar}
\mu=\Phi+\sum_{j=1}^{k-1}\mbox{sgn}(h'_{j})\Phi\circ h_{j}^{-1}.
\end{equation}
The choice of $\Phi$ and $h_{j}$ enables us to construct a map equation with the desired absolutely continuous invariant measure. We can illustrate the use of Eq. (\ref{muvar}) by means of a well-known unimodal map, the tent map
\begin{equation}
\label{tent}
T(x)=1-|2x-1|,
\end{equation}
on $I=[0,1]$. This map was also considered by Lorenz as an approximation for the cusp-shaped Poincar\'e first return map in the Lorenz strange attractor \cite{Lor}. Here we have $h_{1}(x)=h_{1}^{-1}(x)=2-x$ and $\Phi(x)=x/2$, resulting in the uniform invariant density $\rho(x)=1$.

A further advantage that we can draw from such an approach is to obtain conjugate maps. Two measure-preserving maps $S$ and $T$ are thus called if there exists a diffeomorphism $\omega$ such that \cite{CGST,BS}
\begin{equation}
\label{cst}
S\circ\omega=\omega\circ T.
\end{equation}
Conjugacy gathers different maps into a same class where all maps share the same dynamics from the topological viewpoint. More specifically, conjugate maps have the same Lyapunov exponent and their invariant measures are simply related via $\omega$: if $T$ has measure $\mu$, then
\begin{equation}
\label{msw}
\mu_{S}=\mu\circ\omega^{-1}.
\end{equation}
Based on the folding approach above, the transformation $\omega$ relating both maps is such that \cite{TC}
\begin{eqnarray}
\label{conj}
h_{S_{j}}\circ\omega&=&\omega\circ h_{j},\\
\label{pop}
\Phi_{S}&=&\Phi\circ\omega^{-1},
\end{eqnarray}
together with $\epsilon_{j+1}=\epsilon_{j}$ for all $j$, where $\epsilon_{j}=\mbox{sgn}(h'_{j}h'_{Sj})$. Furthermore, $\omega$ is an even function if $\epsilon_{j}=-1$. For example, we can ask for a map $S$ conjugate with tent map (\ref{tent}), let us suppose that $h_{S}(x)=x$. Thus, the transformation is such that $\omega(x)=\omega(2-x)$, with $\epsilon_{1}=-1$. The solution $\omega(x)=\mbox{sn}^{2}(Kx,\kappa)$ fulfills such conditions, where sn is Jacobi's elliptic function, $K(\kappa)$ is the complete elliptic integral of the first kind, and $\kappa$ is its elliptic modulus \cite{AS}. By considering the addition identity of the sn function, this choice gives us Schr\"oder's map \cite{Scp,Ume}
\begin{equation}
\label{schmap}
S_{\kappa}(x)=\frac{4x(1-x)(1-\kappa^{2}x)}{(1-\kappa^{2}x^{2})^{2}},
\end{equation}
with measure $\mu_{S}=\omega^{-1}$. The well-known Ulam's conjugacy $\omega=\sin^{2}(\pi x/2)$ is the $\kappa=0$ particular case of map (\ref{schmap}), resulting in the Ulam-von Neumann logistic map $S_{0}=4x(1-x)$ \cite{Ulam}.

Just for the sake of simplicity, we will focus only on maps with two branches and, therefore, our examples are either unimodal or piecewise expanding maps (see Fig. \ref{fig1}). Special attention will be given to maps having infinite invariant measures. For finite measure cases we also refer the reader to Ref. \cite{TC}.

\section{Infinite measure maps}
\label{sec3}

Let us first consider interval maps, from $[0,1]$ to itself, with a single marginal fixed point at $x=0$:
\begin{equation}
\label{marg}
T(x\rightarrow 0)=0,\qquad T'(x\rightarrow 0)=1.
\end{equation}
The most common form of such maps is
\begin{equation}
\label{mod}
x_{t+1}=f(x_{t})\,\,\,\mbox{mod}\,1,
\end{equation}
with $f:[0,1]\rightarrow[0,2]$ bijective and $h_{1}(x)=x-1$, like the well-known generalization of the Pomeau-Manneville (PM) map \cite{PM}
\begin{equation}
\label{PM}
f(x)=x(1+x^{1/\alpha}),\qquad \alpha>0.
\end{equation}
This kind of map was first obtained from Poincar\'e sections related to the Lorenz attractor for $\alpha=1$ \cite{PM,Lor}.

The remarkable characteristic of such maps is the intermittent switching between long regular phases (so-called laminar) near the indifferent fixed points and short irregular burst ones elsewhere [see Fig. \ref{fig1} (b)]. The applications of such maps have been the most diverse, helping to model anomalous diffusion \cite{GT,GNZ,ZK,DK}, structure of natural languages \cite{EN}, vortex dynamics in evolutive flows \cite{BMI}, weather systems \cite{NEB,RP}, DNA strands of higher eucaryotes \cite{PB}, and neural avalanches in the brain of mammals \cite{ZG}, among others.

One problem to be faced here is the paucity of models whose invariant measures are exact, i.e., analytically known for the whole interval. The most general result that we have so far is the relationship between the invariant density and map equation near the indifferent point $x=0$: $\rho(x)\sim 1/[f'(x)-1]$, up to a multiplicative constant; see Ref. \cite{RV1}.

By considering the approach discussed in Sec. \ref{sec2}, Eq. (\ref{mod}) is such that $T_{0}=f=\Phi^{-1}\circ\mu$, and thus we have from Eq. (\ref{muvar})
\begin{eqnarray}
\label{minf}
\mu(x)&=&\Phi(x)+\Phi(x+1),\\
\label{fapp}
f(x)&=&\Phi^{-1}[\Phi(x)+\Phi(x+1)].
\end{eqnarray}
The choice of $\Phi$ gives us both the map equation and its corresponding invariant measure. Maps with marginal fixed points at $x=0$ are such that $|\Phi'(x\rightarrow0)|\rightarrow\infty$. Thus, from Eq. (\ref{fapp}) one has
\begin{equation}
\label{fmfp}
f(x)\sim x\left[1+\frac{\Phi'(1)}{\Phi'(x)}\right],\qquad x\rightarrow0,
\end{equation}
provided we set $\Phi(1)=0$.

Let us first consider the following choice:
\begin{equation}
\label{d1}
\Phi(x)=\left\{
\begin{array}{ll}
{\displaystyle \frac{x^{1-1/\alpha}-1}{1-1/\alpha}},\quad &\alpha>0,\,\,\alpha\neq1,\\
\\
{\displaystyle  \ln x}, &\alpha=1.
\end{array}
\right.
\end{equation}
For positive $\alpha\neq1$ we have Thaler's map \cite{Thaler}
\begin{equation}
\label{thmap}
f_{\alpha}(x)=x\left[1+\left(\frac{x}{1+x}\right)^{\frac{1-\alpha}{\alpha}}-x^{\frac{1-\alpha}{\alpha}}\right]^{-\frac{\alpha}{1-\alpha}},
\end{equation}
whereas $\alpha=1$ gives us the corresponding PM map (\ref{PM}). Note, by means of Eq. (\ref{fmfp}), that the behavior of Thaler's map near $x=0$ is the same of the PM map (\ref{PM}). For all $\alpha>0$, the invariant density is
\begin{equation}
\label{thinvd}
\rho(x)=b[x^{-1/\alpha}+(1+x)^{-1/\alpha}],
\end{equation}
where $b$ is an undetermined constant for the infinite measure regime $0<\alpha\leq1$, since there is no possible normalization. Interestingly, Eq. (\ref{d1}) is the $1/\alpha$-deformed (or Tsallis \cite{TSA}) logarithm, i.e., $\Phi(x)=\ln_{1/\alpha}x$ for all $\alpha>0$. Thus, both $\alpha=1$ PM and Thaler maps are part of the same expression:
\begin{equation}
\label{adefd}
f_{\alpha}(x)=\exp_{1/\alpha}[\ln_{1/\alpha}x+\ln_{1/\alpha}(x+1)],\qquad\alpha>0,
\end{equation}
with exact invariant density given by Eq. (\ref{thinvd}), where $\exp_{1/\alpha}x$ is the $1/\alpha$-deformed exponential.

The $\alpha=0$ case also gives us a nontrivial class of maps. Two other infinite measure maps are suggested in Ref. \cite{RV1}, and we can also calculate exactly their map equations and corresponding invariant measures for the whole interval. One model is
\begin{equation}
\label{phi2}
\Phi(x)=\exp(x^{-\beta})-e,\qquad \beta>0,
\end{equation}
yielding the map
\begin{equation}
\label{mstr}
f_{\beta}(x)=\ln^{-1/\beta}\{\exp(x^{-\beta})+\exp[(x+1)^{-\beta}]-e\},
\end{equation}
whose behavior near $x=0$ is the same of the map employed in Ref. \cite{DK} to produce strong anomaly diffusion. The other model comes from the choice
\begin{equation}
\label{phi3}
 \Phi_{q,r}(x)=\exp(\ln^{q}x^{r})-1,\qquad r=1-1/\alpha,
\end{equation}
where the parameter $q$ mediates the strong laminar map (\ref{mstr}) as $q\rightarrow\infty$ and Thaler's map at $q=1$. For the choice (\ref{phi3}) we have
\begin{equation}
\label{fmeu}
f_{q,r}(x)=\Phi_{q^{-1},r^{-q}}[1+\Phi_{q,r}(x)+\Phi_{q,r}(x+1)].
\end{equation}
For both cases (\ref{mstr}) and (\ref{fmeu}) the invariant measure is given by direct substitution of corresponding $\Phi$ into Eq. (\ref{minf}).

From the models discussed above, many other maps with exact invariant measures can be generated via topological conjugacy if we just choose a diffeomorphism $\omega$ from $[0,1]$ to itself. For example, $\omega=f/2$ for any $f$ given by Eq. (\ref{fapp}) are possible choices. If we choose $\omega=f_{1/2}/2$ from map (\ref{thmap}), a new rational map $U=\omega\circ T_{1/2}\circ\omega^{-1}$ conjugate with Thaler's map $T_{1/2}$ follows:
\begin{equation}
\label{umap}
U(x)=\left\{
\begin{array}{ll}
{\displaystyle \frac{1}{2}f_{1/2}(2x)=\frac{x(1+2x)}{1+2x-4x^{2}}},\quad &x\in [0,1/2),\\
\\
{\displaystyle  \frac{1}{2}f_{1/2}\circ h_{1}(2x)=\frac{x(1-2x)}{1-6x+4x^{2}}}, &x\in [1/2,1],
\end{array}
\right.
\end{equation}
with measure $\mu_{U}(x)=\mu_{1/2}\circ f_{1/2}^{-1}(2x)$. The map $U$ above has the same behavior of $\alpha=1/2$ in Thaler's map near $x=0$.

A seemingly different case from the indifferent fixed point maps discussed here is Boole's map \cite{AW}
\begin{equation}
B(x)=x-\frac{1}{x},
\end{equation}
from $\mathbb{R}\setminus\{0\}$ to itself. It is easy to see that map $B(x)$ has uniform invariant density throughout a real line, and thus infinite measure, if we look at Boole's integral identity $\int_{-\infty}^{\infty}g(x)dx=\int_{-\infty}^{\infty}g\circ B(x)dx$, valid for any integrable function $g$ \cite{AW}. Despite not having an indifferent fixed point, Boole's map behaves like a $\alpha=1/2$ map. For example, $V=\omega^{-1}\circ B\circ\omega$ with $\omega=1/(1-x)-1/x$ on $(0,1)$ yields the rational map \cite{Thc}
\begin{equation}
\label{vmap}
V(x)=\left\{
\begin{array}{ll}
{\displaystyle \frac{x(1-x)}{1-x-x^{2}}},\quad &x\in [0,1/2),\\
\\
{\displaystyle \frac{1-2x}{1-3x+x^{2}}}, &x\in [1/2,1],
\end{array}
\right.
\end{equation}
with indifferent fixed points $x=0$ and $x=1$, and measure $\mu_{V}=\mu_{B}\circ\omega=\omega$.

\section{Invariant measure as a measurement of information}
\label{sec4}

The exact characterization of an invariant measure is known to be a key issue in nonlinear dynamics because it rules the occupation probabilities of the iterates over entire phase space. Nevertheless, the predictive potential of exact measures goes beyond the statistical description of deterministic systems. We intend to explore here an idea that was recently introduced in Ref. \cite{NV}: the precise knowledge of the measure also enables us to estimate information by observing the transition dynamics along the phase space.

\begin{figure}[ht]
\begin{center}
\resizebox{0.9\linewidth}{!}{\includegraphics*{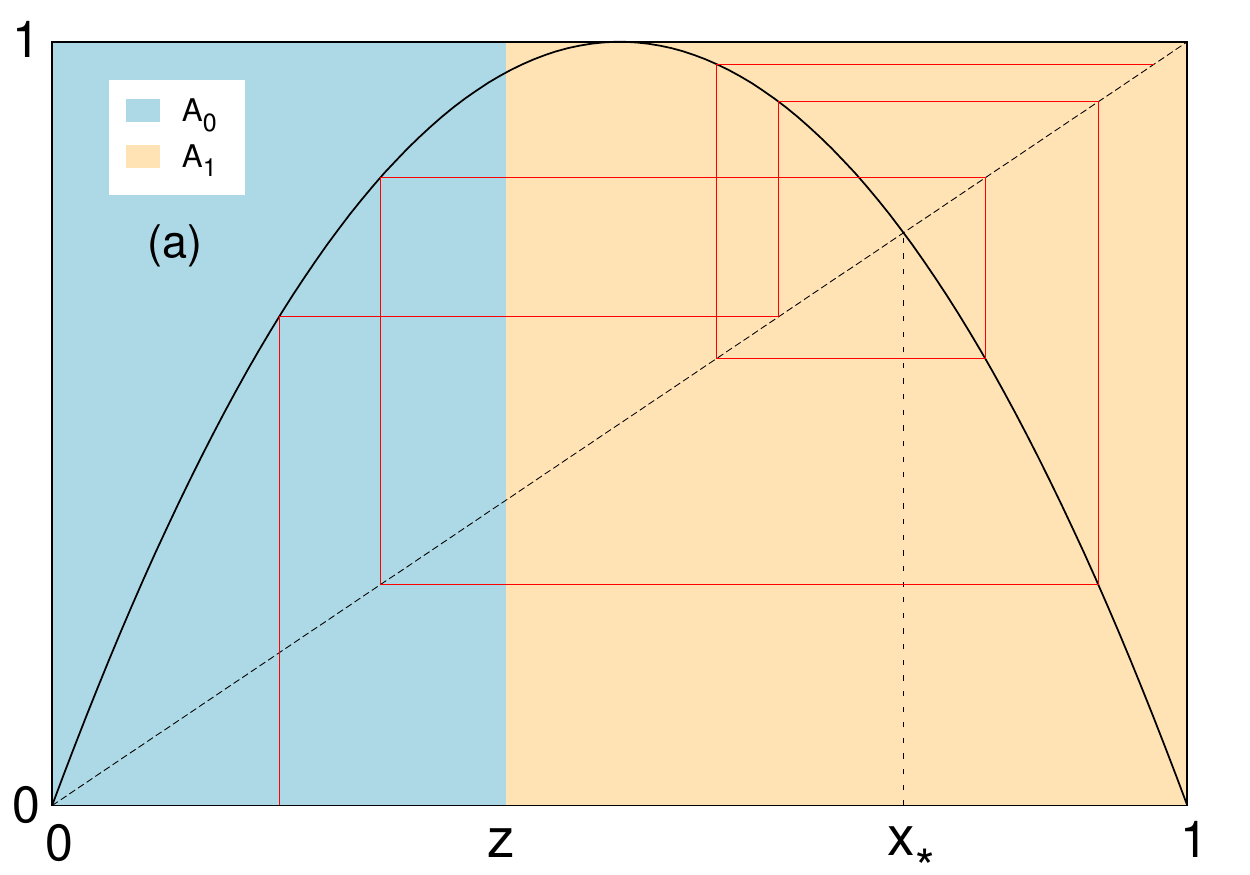}}
\resizebox{0.9\linewidth}{!}{\includegraphics*{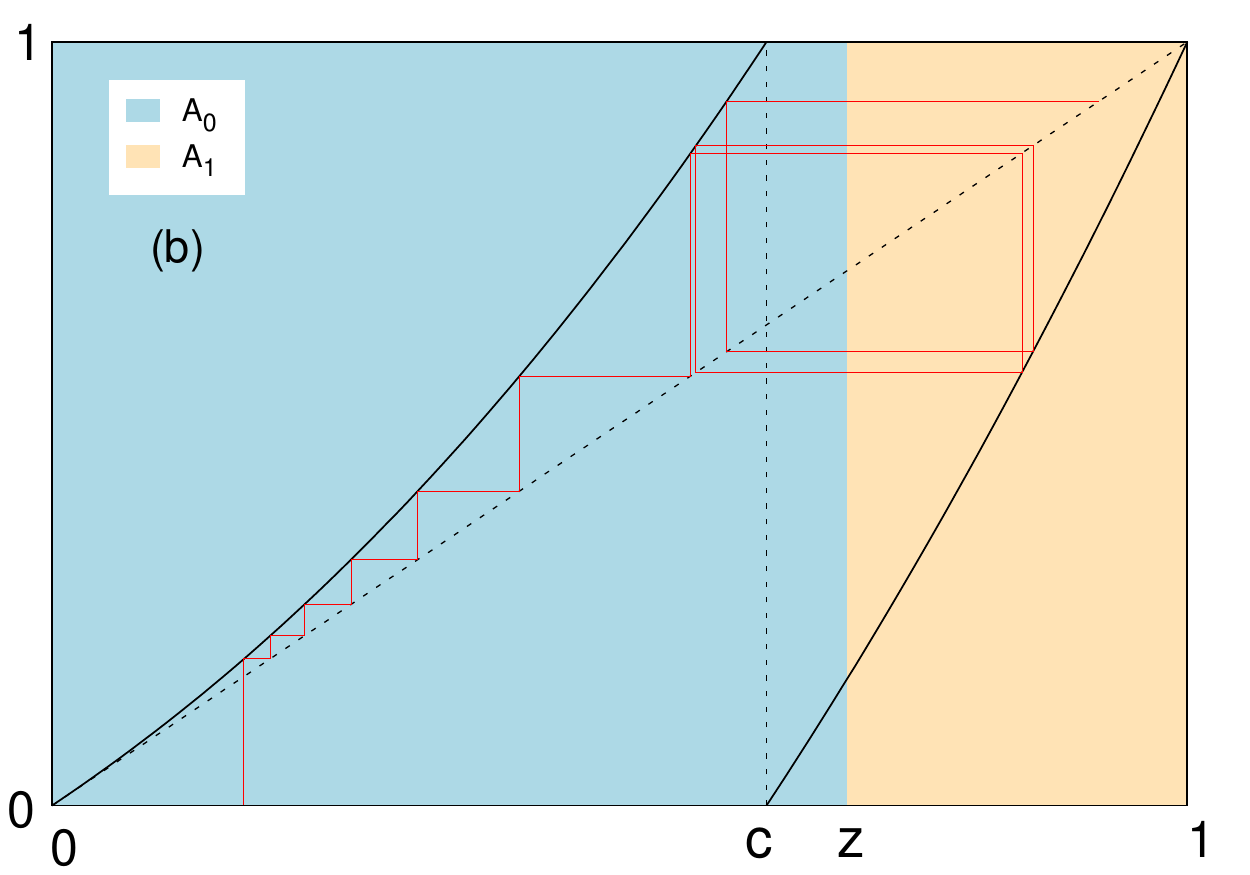}}
\end{center}
\caption{(Color online) Cobweb diagrams for unimodal (a) and piecewise expanding (b) maps. Instead of considering long trajectories as data sets, we translate the dynamics to states $A_{0,1}$. The shape of maps leaves specific features in the number of jumps from $A_{0}$ to $A_{1}$, which can be inferred from exact invariant measures.}
\label{fig1}
\end{figure}

We often turn to symbolic dynamics to simplify analysis of a dynamical system, representing their trajectories by infinite length sequences using a finite number of symbols. Since we are dealing here with two-branches maps, let us consider the usual partition of the interval $[0,1]$ into two cells, $A_{0}=[0,z]$ and $A_{1}=(z,1]$. According to this approach, any trajectory $\{x_{t}\}$ can be represented by a sequence of binary digits $\{s_{t}\}$ such that $s_{t}$ corresponds to the cell where $x_{t}$ belongs, i.e., $s_{t}=k$ for $x_{t}\in A_{k}$, see Fig. \ref{fig1}. Then, redundancies that may appear in $\{s_{t}\}$ are eliminated by considering the algorithmic information, which is defined as the length of the shortest possible program able to reconstruct the sequence $\{s_{t}\}$ on a universal Turing machine \cite{GW}.

Within the symbolic dynamics approach, easier procedures are possible if the invariant measure is exactly known. To better understand how the invariant measure can assist in eliminating redundancies and making the calculation of observables more efficient, let us consider the following observable \cite{NV}:
\begin{equation}
\label{fpc}
\chi(x)=\sigma[T(x)][1-\sigma(x)],
\end{equation}
where $\sigma(x\in A_{1})=1$  and $\sigma(x\in A_{0})=0$. Equation (\ref{fpc}) works like a filter, returning $1$ whenever a transition from $A_{0}$ to $A_{1}$ happens and $0$ otherwise. Now, we can compare the convergence of time averages for different observables under the same trajectories by using the Stepanov-Hopf ratio ergodic theorem \cite{Rcm}
\begin{equation}
\label{sherg}
\frac{\sum_{k=0}^{t-1}\vartheta[T^{k}(x)]}{\sum_{k=0}^{t-1}\varphi[T^{k}(x)]}\rightarrow\frac{\int\vartheta d\mu}{\int\varphi d\mu},
\end{equation}
holding almost everywhere as $t\rightarrow\infty$ for non-negative observables as long as they are integrable over $\mu$. The limiting ratio (\ref{sherg}) can also be regarded as a consequence of Birkhoff's ergodic theorem \cite{Birk} for the particular case of finite measure ergodic maps. On the other hand, in the case of infinite measure ergodic maps \cite{Erg}, we have to deal with averages of the type $\sum_{k=0}^{t-1}\vartheta[T^{k}(x)]/\zeta(t)$, which does not converge to a simple constant. In fact, it strongly depends on the initial condition $x$ (more details in Sec. \ref{sec5}). Thus, the usefulness of ratio (\ref{sherg}) is evident: it gives us pointwise convergence without depending on the finiteness of measure.

By means of filter (\ref{fpc}) we leave aside the complete description of trajectories considering only the jumps from $A_{0}$ to $A_{1}$. From ratio (\ref{sherg}) we have
\begin{equation}
\label{cn}
\sum_{k=0}^{t-1}\vartheta[T^{k}(x)]\sim\gamma N_{t}(x),\qquad \gamma=\frac{\int\vartheta d\mu}{\int\chi d\mu},
\end{equation}
as $t\rightarrow\infty$, where $N_{t}(x)$ is the number of jumps from $A_{0}$ to $A_{1}$, given by
\begin{equation}
\label{Nt}
N_{t}(x)=\sum_{k=0}^{t-1}\chi[T^{k}(x)].
\end{equation}
Thus, the calculation of  $\int\chi d\mu$ plays a central role here since it depends on the partition $z$. For unimodal maps we have (see remark \cite{nrd})
\begin{eqnarray}
\label{siun}
\int\chi d\mu=\left\{
\begin{array}{ll}
  \mu[T_{0}^{-1}(z),z],  & \displaystyle z\leq x_{*}, \\
  \mu[T_{0}^{-1}(z),T_{1}^{-1}(z)], & \displaystyle z>x_{*},
\end{array}
\right.
\end{eqnarray}
where $x_{*}$ is the nonzero root of $x_{*}=T(x_{*})$, whereas for piecewise expanding maps we have
\begin{eqnarray}
\label{sig}
\int\chi d\mu=\left\{
\begin{array}{ll}
  \mu[T_{0}^{-1}(z),z],  & \displaystyle z\leq c, \\
  \mu[T_{0}^{-1}(z),c], & \displaystyle z>c.
\end{array}
\right.
\end{eqnarray}

It is easy to see that, for both Eqs. (\ref{siun}) and (\ref{sig}), the quantity $\int\chi d\mu$ has concave behavior with global maxima at $z=x_{*}$ and $z=c$, respectively. Thus, the coefficient $\gamma$ has a unique global minimum at these partitions, thereby maximizing the number of entrances $N_{t}$. Interestingly, the way how $N_{t}$ is maximized defines whether the map is unimodal or expanding, because $x_{*}$ and $c$ have quite distinct roles in the evolution of trajectories.

One of the applications is the calculation of finite-time Lyapunov exponents $\Lambda_{t}(x)$, which rules the asymptotic separation of initially nearby trajectories according to
\begin{equation}
\label{gr1}
|\delta x_{t}|\sim|\delta x_{0}|\exp[\Lambda_{t}(x_{0})\zeta(t)].
\end{equation}
For finite measure maps, it is widely known that $\zeta(t)\sim t$. On the other hand, for infinite measure maps, the most general instability regime is given by the sublinear growth rate
\begin{equation}
\label{grwz}
\zeta(t)\sim l_{\alpha}(t)t^{\alpha},\qquad 0\leq\alpha\leq1,
\end{equation}
where $l_{\alpha}(t)$ is slowly varying at infinity and with particular asymptotes $l_{0}(t\rightarrow\infty)=\infty$ and $l_{1}(t\rightarrow\infty)=0$. Slowly varying means that $\lim_{x\rightarrow\infty}l_{\alpha}(ax)/l_{\alpha}(x)=1$ for any positive $a$. The direct relationship between map equation and sublinear growth rate (\ref{grwz}) was recently given in Ref. \cite{RV1} (see also Ref. \cite{NV}), namely,
\begin{eqnarray}
\label{s1}
\zeta(t)\sim\frac{1}{\phi^{-1}(t)}\times\left\{
\begin{array}{ll}
  \sin(\pi\alpha)/\pi\alpha,  & \displaystyle 0\leq\alpha<1, \\
  1/\ln t, & \displaystyle \alpha=1,
\end{array}
\right.
\end{eqnarray}
where $T'\sim1-1/x\phi'$ as $x\rightarrow0$. By comparison with Eq. (\ref{fmfp}) we have
\begin{equation}
\label{lan}
\phi'(x)\sim-\frac{1}{x[x/\Phi'(x)]'},\qquad x\rightarrow0,
\end{equation}
up to a multiplicative constant. Evidently, $\phi(x\rightarrow0)$ rules $\phi^{-1}(t\rightarrow\infty)$. The generalized finite-time Lyapunov exponent is therefore given by $\vartheta=\ln|T'|$, but it is much more simple to calculate the number of entrances $N_{t}(x)$ than the summation $\sum_{k=0}^{t-1}\ln|T'(x)|$ if $\mu$ is exactly known. Thus, we have the alternative and more efficient formula
\begin{equation}
\label{lan}
\Lambda_{t}(x)\sim\gamma\frac{N_{t}(x)}{\zeta(t)},\qquad t\rightarrow\infty.
\end{equation}
This approach was successfully employed for Thaler's map; see Ref. \cite{NV}. Of course, Eq. (\ref{lan}) can also be used for finite measure maps and unimodal maps via Eq. (\ref{siun}) with $\zeta(t)=t$.

\section{Infinite measure implies Mittag-Leffler statistics of observables}
\label{sec5}

Given that summations $\sum_{k=0}^{t-1}\vartheta[T^{k}(x)]$ strongly depend on the initial condition $x$ for infinite measure systems, we are interested in the statistics of such quantities when $x$ behaves like a random variable. Following the Darling-Kac approach \cite{DMK}, the strategy for finding the asymptotic statistics is based on the calculation of moments,
\begin{equation}
\label{ik}
I_{n}(t)=E\left\{\left[\sum_{k=0}^{t-1}\vartheta(T^{k}(x))\right]^{n}\right\},
\end{equation}
with $E\{\,\,\}$ denoting the expectation value.

The limiting ratio (\ref{sherg}) tells us that the asymptotic statistics of $I_{n}(t)$ does not depend on the choice of non-negative observable $\vartheta$ provided it is integrable over $\mu$ and that the summation $\sum_{k=0}^{t-1}\vartheta[T^{k}(x)]$ is suitable normalized. Furthermore, ratio (\ref{sherg}) also suggests the existence of an asymptotically separable density function $p(x,t)\equiv p_{t}(x)\sim r_{t}\rho(x)\sim r(t)\rho(x)$ such that
\begin{eqnarray}
\label{say}
I_{1}(t)&=&\sum_{k=0}^{t-1}\int\vartheta(x)p_{k}(x)dx\sim\sum_{k=0}^{t-1}r_{k}\int\vartheta(x)\rho(x)dx\nonumber\\
&\sim&\int r(t)dt\int\vartheta d\mu
\end{eqnarray}
as $t\rightarrow\infty$, where $r(t)$ does not depend on the observable, so that $\int r(t)dt$ cancels itself out. For Lyapunov exponents ($\vartheta=\ln|T'|$) we know that $r(t)\propto\zeta'(t)$ due to subexponential instability, therefore $p(x,t)\sim\zeta'(t)\rho(x)$ as $t\rightarrow\infty$, up to a multiplicative constant. This asymptotic separable form also proves to be quantitatively consistent with relationships between infinite measure and subexponential instability; see Ref. \cite{RV1}.

If, on the one hand, the limiting ratio (\ref{sherg}) tells us that the statistics of observables is universal, on the other hand, Darling-Kac approach requires the use of a Markov process with stationary transitions. Given that $T$ is a Markov map, the choice of filter (30) via standard partition of the phase space meets this criterion since most part of laminar trajectories are compressed as ``state $A_{0}$''.

By performing the Laplace transform $t\mapsto s$ on Eq. (\ref{say}) with $\vartheta=\chi$, the first moment is given by
\begin{equation}
\label{c1}
\frac{\tilde{I}_{1}(s)}{\tilde{\zeta}(s)}=\int\frac{\tilde{p}(x,s)}{s\tilde{\zeta}(s)}\chi(x)dx\rightarrow C,\qquad s\rightarrow0,
\end{equation}
being $C\propto\int\chi d\mu$ a positive constant. The calculation of the second moment is as follows:
\begin{equation}
\label{c2}
I_{2}(t)\sim2!\int\int\chi(x_{1})\chi(x_{2})y_{2}(x_{1},x_{2},t)dx_{1}dx_{2},
\end{equation}
where
\begin{equation}
\label{c3}
y_{2}(x_{1},x_{2},t)=\int_{0}^{t}\int_{0}^{t_{2}}p(x_{1},t_{1})p(x_{2},t_{2}-t_{1})dt_{1}dt_{2},
\end{equation}
resulting in $s\tilde{I}_{2}(s)[s\tilde{\zeta}(s)]^{2}\rightarrow 2!C^{2}$ as $s\rightarrow0$. Similar calculations lead to the immediate extension for all $n\geq1$:
\begin{equation}
\label{i2s}
s\frac{\tilde{I}_{n}(s)}{[s\tilde{\zeta}(s)]^{n}}\rightarrow n!C^{n},\qquad s\rightarrow0.
\end{equation}
Now, by applying Karamata's Tauberian theorem \cite{Feller} to Eqs. (\ref{grwz}) and (\ref{i2s}), one finally has
\begin{equation}
\label{MLm}
\lim_{t\rightarrow\infty}E\left\{\left[\frac{\sum_{k=0}^{t-1}\chi(T^{k}(x))}{C\zeta(t)}\right]^{n}\right\}=n!\frac{\Gamma^{n}(1+\alpha)}{\Gamma(1+\alpha n)}.
\end{equation}

For $\alpha=1$ we have the degenerate case
\begin{equation}
\label{wet}
\frac{\sum_{k=0}^{t-1}\chi(T^{k}(x))}{\zeta(t)}\rightarrow C,\qquad t\rightarrow\infty,
\end{equation}
which is similar to Birkhoff's ergodic theorem for linear $\zeta(t)$ \cite{Birk}. For $0\leq\alpha<1$, the numbers on the right side of Eq. (\ref{MLm}) are just the moments of the Mittag-Leffler probability distribution with unit expected value. For $\alpha=0$ the Mittag-Leffler distribution becomes the exponential distribution $1-e^{-\xi}$ on $\xi\geq0$. For $0<\alpha<1$ the corresponding probability density function $m_{\alpha}(\xi)$ is
\begin{equation}
\label{dML}
m_{\alpha}(\xi)=\frac{\Gamma^{1/\alpha}(1+\alpha)}{\alpha \xi^{1+1/\alpha}}\,g_{\alpha}\left[\frac{\Gamma^{1/\alpha}(1+\alpha)}{\xi^{1/\alpha}}\right],
\end{equation}
where $g_{\alpha}$ stands for the one-sided L\'evy stable density, whose Laplace transform is $\tilde{g}_{\alpha}(s)=\exp(-s^{\alpha})$ \cite{ML}. Since the Mittag-Leffler distribution is uniquely characterized by its moments (\ref{MLm}), one has  
\begin{equation}
\label{ADK}
\lim_{t\rightarrow\infty}\mbox{Prob}\left\{\frac{\sum_{k=0}^{t-1}\chi(T^{k}(x))}{C\zeta(t)}<z\right\}=\int_{0}^{z}m_{\alpha}(\xi)d\xi.
\end{equation}
The constant $C$ is given by $C=\kappa\int\chi d\mu$, where $\kappa$ is such that $C$ is well defined even in the infinite measure regimes; see Ref. \cite{NV} for more details. The quantity $C\zeta(t)$ is known as a {\it return sequence} in the infinite ergodic theory \cite{Aaronson}.

The results above give the statistics of the number of jumps $N_{t}$ but, as we have previously mentioned, such results are also valid for any non-negative observables that are integrable over $\mu$; see Eq. (\ref{cn}). It is worth mentioning that the statistics of $N_{t}$ is usually considered in the literature by means of renewal theory. Despite providing the correct Mittag-Leffler statistics of jumps, renewal theory does not provide the real number of entrances into $A_{1}$; for a detailed discussion see Ref. \cite{NV}.

Another important point is that the Darling-Kac approach also tells us that, inversely, Mittag-Leffler statistics implies sublinear growth rate (\ref{grwz}); see Ref. \cite{DMK} for more details.

\section{Concluding remarks}

The goal of this paper was to investigate for which kinds of mappings there exist invariant measures, how many of them, whether they are equivalent (from the topological viewpoint), and what are their ergodic properties. Typically, these issues are discussed in the framework of finite measure systems, separately from the infinite ones, which are relatively less known. Here we took a different perspective: by using a new IPFP approach, invariant measures are placed as the starting point, and their connections with the kind of instability (weakly or fully chaotic) and corresponding ergodic regimes are then drawn. For one-dimensional ergodic maps, Mittag-Leffler statistics of observables is an unequivocal signature of subexponential Lyapunov instability (weak chaos), and this is a direct consequence of infinite invariant measures. After some parametric change, e.g., from $\alpha<1$ to $\alpha>1$, if an infinite invariant measure becomes a probability measure, then Lyapunov instability becomes exponential and Mittag-Leffler density becomes a Dirac $\delta$ function, which are signatures of finite measure ergodic systems. Thus, the strength of invariant measure plays the role of order parameter between weak (subexponential) and fully developed chaos.

Subexponential instability implies zero Lyapunov exponent in the conventional sense, and its relationship with infinite measures, as outlined here, may have parallels to other kinds of systems. We can mention, for instance, chaotic attractors that lose stability in some invariant subspace by varying a parameter. For such systems, the so-called blowout bifurcation \cite{OJS} occurs when the most positive normal Lyapunov exponent of the attractor crosses zero at the point of loss of stability. Interestingly, at least two models exhibit infinite measures similar to $\alpha=1$ PM maps when undergoing blowout bifurcation: chaotic motion of two identical dissipatively coupled one-dimensional mappings \cite{PGRS} and the drift-diffusion model \cite{AAN}.

\begin{acknowledgements}
The author thanks Pierre Naz\'e for helpful discussions. This work was supported by Conselho Nacional de Desenvolvimento Cient\'ifico e Tecnol\'ogico (CNPq), Brazil (Grant No 307618/2012-9) and special program PROPES-Multicentro
(UFABC), Brazil.
\end{acknowledgements}


\begin{thebibliography}{99}
\bibitem{nrd} According to the well-known Radon-Nikodym theorem, an absolutely continuous invariant measure $\mu$ with respect to the Lebesgue measure on set $A$ is such that $\mu(A)=\int_{A}\rho(x)dx$. From our notation this is equivalent to $\mu([a,x])\equiv\mu(x)-\mu(a)$. Thus, $\mu(x)$ denotes de invariant measure calculated on an arbitrary interval up an additive (unimportant) constant.
\bibitem{Erg} A map $T$ is ergodic if $\mu(A)=0$ or $\mu(A^{c})=0$ for any invariant set $A=T^{-1}(A)$, where $A^{c}$ is the complement of $A$. This definition also holds for infinite measure maps.
\end{thebibliography}
\end{document}